\documentclass[aps,prl,twocolumn,showpacs]{revtex4}
\usepackage{amssymb}
\usepackage{amsfonts}
\usepackage{amsmath}
\usepackage{mathrsfs}

\begin{document}

\title{Poincar\'e subalgebra and gauge invariance in nucleon structure}

\author{Xiang-Song Chen}
\email{cxs@hust.edu.cn}

\affiliation{Department of Physics, Huazhong
University of Science and Technology, Wuhan 430074, China}
\affiliation{Joint Center for Particle, Nuclear Physics and
Cosmology, Nanjing 210093, China}
\affiliation{Kavli Institute for Theoretical Physics China, Chinese
Academy of Science, Beijing 100190, China}

\date{\today}

\begin{abstract}
By separating the gluon field into physical and pure-gauge
components, the usual Poincar\'e subalgebra for an
interacting system can be reconciled with gauge-invariance when
decomposing the total rotation and translation generators
of QCD into quark and gluon parts. The gauge-invariant quark/gluon
parts act as the generators for the gauge-invariant physical
component of the quark/gluon field, not the full
quark/gluon field which also contains the gauge degrees of freedom.
We clarify that the naive canonical decomposition of generators,
while trivially respecting the Poincar\'e subalgebra, might not
give a completely gauge-invariant quark-gluon structure of
the nucleon momentum and spin, though limited invariance within
a certain gauge class can be proven.

%14.20.Dh Protons and neutrons
%11.15.-q Gauge field theories
%12.38.-t Quantum chromodynamics
\end{abstract}
\maketitle

{\it Introduction.}---In the past few years, the nucleon spin
problem \cite{Spin} received an increasing theoretical devotion,
which, however, has amazingly caused more controversy in analyzing
the nucleon spin in terms of the quark and gluon contributions.
In earlier study of the nucleon spin ``crisis'' that quarks do not
polarize significantly inside the nucleon,
a major difficulty for about two decades had been the entanglement
of gauge-invariance with the most natural idea of seeking the
``missing'' nucleon spin
from the gluon polarization or the orbital angular momentum of
quarks and gluons. A turning point was brought by
Refs.~\cite{Chen08,Chen09}, which employ the idea of decomposing
the gluon field into physical and pure-gauge components, thus offer
much freedom in constructing gauge-invariant quantities. But soon,
Wakamatsu \cite{Waka} and Leader \cite{Lead11} argue that the
freedom becomes too much that the
gauge-invariant decomposition of the nucleon spin into spin and
orbital contributions of quarks and gluons is actually not unique.
To clarify the issue, Leader advocates a compelling
theoretical criteria that the angular momentum operator of the quark/gluon
field should generate spatial rotation for the quark/gluon field,
respectively (and similarly for the linear momentum operator) \cite{Lead11}.
Then Leader concludes that such a criteria
refers unambiguously to the naive, canonical decomposition of the QCD rotation
generator as originally discussed by Jaffe and Manohar \cite{Jaff90},
in which the quark/gluon part takes the ``free'' form
as if the quark/gluon field were existing alone (A quotation mark is put
on ``free'' because the gluon field is still subjected to self-interaction).
Such free-form operators are naturally gauge-dependent,
but Leader argues that in quantum theory
what matters is not the operator itself, but the matrix element of the
operator, and that these free-form angular momentum operators
do give gauge-invariant expectation values in the nucleon helicity eigenstate.
In this regard, it should be noted that the idea of gauge-invariant
matrix elements of some gauge-dependent operators in certain physical states
had been put forward as early as in 1995 by Anselmino, Efremov, and Leader
in Ref. \cite{Anse95}, and later had been seriously pursued by the present 
and other authors, including a formal supportive proof
with the path-integral approach \cite{Chen98}, explicit perturbative calculation
with opposite conclusion \cite{Hood99}, and also a revealing of the
unreliability of the utilized conventional path-integral approach \cite{Chen99,Sun00}.

This paper aims to reduce (and hopefully remove) the controversy, by presenting
the real unique answer dictated by Poincar\'e subalgebra and gauge invariance.

{\it Poincar\'e subalgebra for an interacting system.}---It is fair
to say that if without gauge symmetry, the problem of a schematic
separation of the angular momentum for an interacting
system would be fairly trivial. Let us recall the well-known
structure of the ten Poincar\'e generators for an interacting system of
two fields, $\phi_E$ and $\phi_F$, which we collectively denote as
$\phi_X$ (with $X=E,F$).
In the instant form, six generators are
interaction-free (or {\em good}): $\vec P=\vec P_E+\vec P_F$, $\vec
J=\vec J_E +\vec J_F$, and four are interaction-involving (or {\em bad}):
$H=H_E+H_F+H_{int}$, $\vec K=\vec K_E+\vec K_F+\vec K_{int}$, where
$int$ denotes the interacting part.
In the interaction-free generators, $\vec P_X$ and $\vec J_X$
satisfy their own subalgebra:
\begin{equation}
\left[ P^i_X, P^j_X \right]=0,
\left[ J^i_X, J^j_X \right]=i\epsilon_{ijk} J^k_X,
\left[ J^i_X, P^j_X \right]=i\epsilon_{ijk} P^k_X.
\label{algebra}
\end{equation}

For a gauge-interaction system, however, such subalgebra alone does not prescribe
a unique separation of $\vec J$ and $\vec P$. To unambiguously pin down the
separation, Leader advocates that $\vec J_X$ and $\vec P_X$
should truly act as the rotation and translation generators of
$\phi_X$. Because $\vec J_X$ and $\vec P_X$ are not conserved separately,
the operation has to be at the same instant \cite{Lead11}:
\begin{subequations}
\label{generator}
\begin{eqnarray}
i\left[ \vec P_X(t), \phi_X(\vec x, t) \right]=
\vec\partial \phi_X(\vec x, t),\\
i\left[\vec J_X (t), \phi_X(\vec x, t) \right ]=
(\vec x \times \vec \partial +\vec {\mathcal S}_X ) \phi_X(\vec x, t),
\end{eqnarray}
\end{subequations}
where ${\mathcal S}_X $ is the spin matrix that governs the Lorentz
transformation of $\phi_X$. For a scalar field we have of course
$ {\mathcal S}_X =0$

Eqs. (\ref{generator}) are much stronger requirements than Eqs. (\ref{algebra}),
which are in fact the corollary of Eqs. (\ref{generator}).
Leader concludes that Eqs. (\ref{generator}) select unambiguously the
naive canonical decomposition of the total $\vec J$ and $\vec P$ of QCD
as originally discussed by Jaffe and Manohar \cite{Jaff90}:
\begin{subequations}
\label{nc}
\begin{eqnarray}
\vec P &=& \int d^3x \psi ^\dagger \frac 1i \vec \partial  \psi
+\int d^3x E^i \vec \partial A^i \nonumber \\
&\equiv& \vec P^{nc}_q + \vec P^{nc}_g \label{Pnc},\\
\vec J &=& \int d^3 x \psi ^\dagger
(\frac 12 \vec \Sigma +\vec x \times\frac 1i \vec \partial)\psi
\nonumber \\
 &+&\int d^3x (\vec E \times \vec A+
E^i\vec x\times \vec \partial A^i) \nonumber \\
&\equiv& \vec J^{nc}_q +\vec J^{nc}_g, \label{Jnc}
\end{eqnarray}
\end{subequations}
Here ``$nc$'' means ``naive canonical'', in the sense that the operators
$\vec P^{nc}_{q/g}$ and $\vec J^{nc}_{q/g}$ take their free-form
expressions in a canonical formulation. Such free-form operators naturally
respect Eqs. (\ref{generator}) (with $\phi_X=\psi, \vec A$),
but are also naturally gauge-dependent. The reconciliation of Eqs. (\ref{generator})
with gauge-invariance, however, turns out to be extremely troublesome.
For example, the widely employed gauge-invariant decomposition \cite{Ji97},
\begin{subequations}
\label{gi}
\begin{eqnarray}
\vec P &=& \int d^3x \psi ^\dagger \frac 1i \vec D \psi
+\int d^3x \vec E \times \vec B \nonumber \\
&\equiv& \vec P_q^{gi} + \vec P_g^{gi}, \label{Pgi} \\
\vec J&=& \int d^3 x \psi ^\dagger
(\frac 12 \vec \Sigma + \vec x \times\frac 1i \vec D) \psi %\nonumber\\&&
+\int d^3x \vec x\times (\vec E \times\vec B ) \nonumber \\
&\equiv &\vec J_q^{gi} +\vec J_g^{gi},  \label{Jgi}
\end{eqnarray}
\end{subequations}
where $\vec D=\vec\partial-ig\vec A$ is the gauge-covariant derivative
and ``$gi$'' denotes ``gauge-invariant'',
evidently do not respect Eqs. (\ref{generator}). In fact,
this explicitly gauge-invariant decomposition does not even manifest
the interaction-free feature of $\vec P$ and $\vec J$, and violates the
subalgebra in Eqs. (\ref{algebra}). (While the coupling term in
$\vec P_q^{gi}$ can be removed by choosing a gauge, that in $\vec J_q^{gi}$
is substantial and cannot.)

Since Eqs. (\ref{generator}) comprise compelling criteria in defining
momentum and angular momentum, Leader advocates that a pertinent
analysis of the quark-gluon structure of the nucleon momentum and spin
should be based on Eqs. (\ref{nc}), not Eqs. (\ref{gi}) or any other
proposals in the literature.
As to the gauge-dependence problem with Eqs. (\ref{nc}), Leader claims
that it is unsubstantial, and proves in Ref. \cite{Lead11}
that the relevant observables, namely the expectation values of the
operators $\vec P^{nc}_q$, $\vec J^{nc}_q$ etc. in a nucleon helicity
eigenstate, are nevertheless gauge-invariant.

{\it Gauge invariance of operator and matrix element.}---Over ten year ago,
we had been attracted by exactly the same idea as Leader's that some
gauge-dependent operators may produce gauge-invariant matrix elements in
certain physical states. If applicable to the operators in Eqs. (\ref{nc}),
this idea would greatly simplify the gauge-invariance problem in nucleon structure.
But unfortunately, Leader's discussion in Ref. \cite{Lead11} was fully based
on covariant quantization, hence what Leader proved is only a partial
invariance within the covariant gauge. In a never published preprint \cite{Chen98},
a proof was given for general gauges, utilizing the standard path-integral formalism.
Explicit perturbative calculation was carried out in Ref. \cite{Hood99}, which
confirms the invariance within the covariant gauge, but shows distinct results for
the covariant and light-cone gauges. The conflict between formal path-integral
proof and explicit perturbative calculation led to serious questioning of the
reliability of the path-integral approach in a gauge theory. In Ref. \cite{Chen99}, it
was show that the standard path-integral formulation can be used to prove that
the fermion two-point Green function in Abelian theory is gauge invariant, which is
evidently incorrect. Ref. \cite{Sun00} demonstrated further that the commonly employed
procedures such as averaging over the gauge group and interchanging the integration
order might also lead to incorrect conclusions.

Considering all these troubles, we finally gave up the very attractive idea that
the naive canonical decomposition in Eqs. (\ref{nc}) might 
prescribe a gauge-invariant nucleon structure. It should be remarked that 
the use of gauge-dependent operators is after all unsafe and
compromising, since they cannot possibly guarantee gauge-invariance of {\it all}
matrix elements (otherwise the operator is gauge-invariant by definition).
In the following, we will carefully demonstrate that one does
not have to compromise so much on such a fundamental principle as
gauge-invariance. In fact, it is possible to reconcile the nontrivial
Eqs. (\ref{generator}) with gauge-invariance {\em at the  operator level},
which then safely guarantees gauge-invariance of the matrix elements, and
hence a gauge-invariant, physically meaningful quark-gluon
decomposition of the nucleon momentum and angular momentum.

{\it Generators for the physical fields.}---As a preparing step, we first
note that the less demanding corollary of Eqs. (\ref{generator}), namely
the Poincar\'e subalgebra in Eqs. (\ref{algebra}), has already been
reconciled with gauge-invariance in Refs. \cite{Chen08,Chen09}. The technique
is to separate the gauge field $A^\mu (x)\equiv A^\mu_{phys}(x)+ A^\mu_{pure}(x)$.
The physical component $A^\mu_{phys}$ has the same gauge-transformation
behavior as the field strength $F^{\mu\nu}$ (namely, $A^\mu_{phys}$ is
gauge-invariant/covariant in Abelian/non-Abelian gauge theory); and the pure-gauge
component $A^\mu_{pure}$ gives null field strength and has the same
gauge-transformation behavior as the full $A^\mu $. Such a separation
brings a great advantage of {\it minimally} upgrading a gauge-dependent expression
to be gauge-invariant. Namely, to seek a gauge-invariant/covariant replacement
for $A^\mu $ one can now use $A^\mu_{phys}$ instead of $F^{\mu\nu}$, and
$A^\mu_{pure}$ can be used instead of $A^\mu$ to construct a gauge-covariant derivative.
This type of upgrading is minimal because $A^\mu_{pure}$ is a pure-gauge, thus can be
removed by gauge-transformation. We will name the particular gauge with $A^\mu_{pure}=0$
the ``physical gauge'', in which the minimally upgraded gauge-invariant expression
reduces to the original gauge-dependent expression.

By the above technique, a gauge-invariant
separation of the rotation and translation generators in gauge theories can be achived,
while respecting the Poincar\'e subalgebra in Eqs. (\ref{algebra}).
For the simpler Abelian case like an electron-photon ($e$-$\gamma$) system,
the explicit expressions are \cite{Chen08}:
\begin{subequations}
\label{gic}
\begin{eqnarray}
\vec P &=& \int d^3x \psi ^\dagger \frac 1i \vec D_{pure}  \psi
+\int d^3x E ^i \vec \partial A_{phys}^i \nonumber \\
&\equiv& \vec P^{gic}_e + \vec P^{gic}_\gamma \label{Pgic},\\
\vec J &=& \int d^3 x \psi ^\dagger
(\frac 12 \vec \Sigma +\vec x \times\frac 1i \vec D_{pure})\psi
\nonumber \\
 &+&\int d^3x (\vec E \times \vec A_{phys}+
E ^i\vec x\times \vec \partial A_{phys} ^i) \nonumber \\
&\equiv& \vec J^{gic}_e +\vec J^{gic}_\gamma. \label{Jgic}
\end{eqnarray}
\end{subequations}
Here $\vec D_{pure}=\vec \partial -iq \vec A_{pure}$ is the pure-gauge
covariant derivative, $q$ is the electron charge, and $gic$ means
``gauge-invariant canonical''. The gauge-invariant operators
$\vec P^{gic}_{e/\gamma}$, $\vec J^{gic}_{e/\gamma}$ are evidently not
the translation and rotation generators for $\psi$ and $\vec A$ which
contain gauge degrees of freedom. But since in the physical gauge with
$\vec A_{pure}=0$ we have $\vec A_{phys}=\vec A$,
a comparison with Eqs. (\ref{nc}) reveals that $\vec P^{gic}_{\gamma}$ and
$\vec J^{gic}_{\gamma}$ are the translation and rotation generators for
the {\em gauge-invariant physical} photon field $\vec A_{phys}$. We now
demonstrate a key point that the electron operators $\vec P^{gic}_e$ and
$\vec J^{gic}_e$ can be rewritten as:
 \begin{subequations}
\label{gic'}
\begin{eqnarray}
\vec P^{gic}_e &=&
\int d^3x \psi_{phys} ^\dagger \frac 1i \vec \partial  \psi_{phys}, \label{Pgic'}\\
\vec J^{gic}_e &=&  \int d^3 x \psi_{phys} ^\dagger
(\frac 12 \vec \Sigma +\vec x \times\frac 1i \vec \partial)\psi_{phys},\label{Jgic'}
\end{eqnarray}
\end{subequations}
with $\psi_{phys}$ a gauge-invariant quantity. To see this, we put
$\psi_{phys}=  e^{-iq\Lambda } \psi$,  then Eqs. (\ref{gic'})
require $\psi^\dagger \vec D_{pure}\psi =\psi^\dagger _{phys}\vec \partial \psi_{phys}$,
which gives $\vec\partial \Lambda = \vec A_{pure}$, or $\Lambda =\frac 1{\vec \partial^2}
\vec\partial\cdot \vec A_{pure}$. So we find
\begin{equation}
\psi_{phys}=\exp{\{-iq\frac 1{\vec \partial ^2}
\vec\partial \cdot \vec A_{pure}\}} \psi . \label{ephys}
\end{equation}
This is evidently invariant under the combined
gauge-transformation $\psi \to e^{ iq\omega }$,
$\vec A\to \vec A+\vec\partial \omega$ (note that $\vec A _{pure}$ and $\vec A$
transform in the sane way). In the physical gauge we have $\psi_{phys}=\psi$,
thus $\vec P^{gic}_e$ and $\vec J^{gic}_e$ are the translation and rotation
generators for the gauge-invariant physical electron field $\psi_{phys}$.
We therefore see that Eqs. (\ref{gic}) also fulfil Eqs. (\ref{generator}),
with $\phi_X=\psi_{phys}, \vec A_{phys}$. Namely, {\it $\vec P^{gic}_{e/\gamma}$
and $\vec J^{gic}_{e/\gamma}$ qualify as the momentum and angular momentum
operators for the gauge-invariant physical electron/photon field.}

For the non-Abelian quark-gluon system, the counterpart of Eqs. (\ref{gic}) is
\cite{Chen09}:
\begin{subequations}
\label{gic-nA}
\begin{eqnarray}
\vec P &=& \int d^3x \psi ^\dagger \frac 1i \vec D_{pure}  \psi
+\int d^3x E ^i \vec {\cal D}_{pure}  A_{phys}^i \nonumber \\
&\equiv& \vec P^{gic}_q + \vec P^{gic}_g \label{Pgic-nA},\\
\vec J &=& \int d^3 x \psi ^\dagger
(\frac 12 \vec \Sigma +\vec x \times\frac 1i \vec D_{pure})\psi
\nonumber \\
 &+&\int d^3x (\vec E \times \vec A_{phys}+
E ^i\vec x\times \vec {\cal D}_{pure} A_{phys} ^i) \nonumber \\
&\equiv& \vec J^{gic}_q +\vec J^{gic}_g. \label{Jgic-nA}
\end{eqnarray}
\end{subequations}
A critical difference from the Abelian case is that the physical gluon
field $\vec A_{phys}$ is now gauge-covariant instead of gauge-invariant,
hence $\vec A_{phys}$ also needs a pure-gauge covariant derivative
$\vec {\cal D}_{pure}=\vec\partial -ig[\vec A_{pure},~] $, so as to make
$\vec P^{gic}_g$ and $\vec J^{gic}_g$ gauge-invariant and at the same
time satisfy the subalgebra in Eqs. (\ref{algebra}). The quark sector,
on the other hand, looks similar to that in Eqs. (\ref{gic}), so we write
analogously to Eqs. (\ref{gic'}),
\begin{subequations}
\label{q-gic-nA'}
\begin{eqnarray}
\vec P^{gic}_q &=& \int d^3x \hat \psi_{phys} ^\dagger \frac 1i \vec \partial
 \hat \psi_{phys}, \label{q-Pgic-nA'}\\
\vec J^{gic}_q &=&  \int d^3 x \hat \psi_{phys} ^\dagger
(\frac 12 \vec \Sigma +\vec x \times\frac 1i \vec \partial)\hat \psi_{phys},
\label{q-Jgic-nA'}
\end{eqnarray}
\end{subequations}
with $\hat \psi_{phys}$ also intended to be gauge-invariant.
We put a hat to distinguish
it from Eq. (\ref{ephys}), which no longer gives a gauge-invariant quantity as
$\vec A_{pure}$ undergoes a non-Abelian transformation. To seek the expression
of $\hat \psi_{phys}$, we again write
$\hat \psi_{phys}=e^{-ig\Lambda } \psi $. But now $\Lambda$ is a matrix:
$\Lambda \equiv \Lambda ^a T^a$, with $T^a$ the generators of the color SU(3) group.
By requiring $\psi^\dagger \vec D_{pure}\psi =\hat \psi_{phys} ^\dagger \vec \partial
 \hat \psi_{phys}$, we get
\begin{equation}
e^{ig\Lambda } \vec\partial e^{-ig\Lambda } =-ig \vec A_{pure}. \label{Lambda}
\end{equation}

When perturbative expansion is allowed,  $\Lambda$ can be solved
uniquely in terms of $\vec A_{pure}$, which in tern is uniquely given by $\vec A$
\cite{Chen09,Chen11d,Chen11p}.
For $\hat \psi_{phys}$ to be gauge-invariant, $e ^{-ig\Lambda }$ must transform to
$e^{-ig\Lambda ' } = e^{-ig\Lambda } U^\dagger$
under the combined gauge transformation: $\psi \to \psi'= U\psi$, $\vec A\to
\vec A'=U\vec A U^\dagger +\frac ig U\vec \partial U^\dagger$. This can be verified
as follows: A slight algebra can show that $e^{-ig\Lambda ' } =
e^{-ig\Lambda } U^\dagger$ is indeed a solution of Eq. (\ref{Lambda}) for
$\vec A'_{pure} = U\vec A'_{pure} U^\dagger +\frac ig U\vec \partial U^\dagger$
(note again that $\vec A_{pure}$ and $\vec A$ transform in the same way); it is then
the unique solution given validity of the perturbative expansion (which we always
assume in this paper by restricting our discussion to the region with small coupling
constant or small field amplitude).

Since $\hat \psi_{phys}$ is gauge-invariant and reduce to $\psi$ in the physical gauge,
$\vec P_q^{gic}$ and $\vec J_q^{gic}$ are the translation and rotation generators for
$\hat \psi_{phys}$.

The gluon sector is far more tricky. The covariant derivative on $\vec A_{phys}$
renders that the gauge-invariant $\vec P^{gic}_g$ and $\vec J^{gic}_g$ are
{\it not} generators for the gauge-covariant $\vec A_{phys}$. After a
careful manipulation, we find that $\vec P^{gic}_g$ and $\vec J^{gic}_g$
can be converted into
\begin{subequations}
\label{g-gic-nA'}
\begin{eqnarray}
\vec P^{gic}_g &=&\int d^3x \hat E_{phys} ^i \vec \partial  \hat A_{phys}^i \label{g-Pgic-nA'},\\
\vec J^{gic}_g &=&  \int d^3x (\vec {\hat E}_{phys} \times \vec {\hat  A}_{phys}
+\hat E_{phys} ^i\vec x\times \vec \partial  \hat A_{phys} ^i). \nonumber\\
\label{g-Jgic-nA'}
\end{eqnarray}
\end{subequations}
Here $\vec {\hat E}_{phys}$, $\vec {\hat A}_{phys}$ are defined as
\begin{equation}
\vec {\hat E}_{phys} =e^{-ig\Lambda} \vec E e^{ig\Lambda }, ~
\vec {\hat A}_{phys} =e^{-ig\Lambda} \vec A_{phys} e^{ig\Lambda}, \label{EA}
\end{equation}
with $\Lambda$ given by Eq. (\ref{Lambda}). $\vec {\hat E}_{phys}$,
$\vec {\hat A}_{phys}$ are evidently gauge-invariant by noting that
$\vec E\to U\vec E U^\dagger$, $\vec A_{phys}\to U\vec A_{phys} U^\dagger$,
and $e^{-ig\Lambda } \to e^{-ig\Lambda } U^\dagger$,
$e^{ig\Lambda } \to U e^{ig\Lambda } $. The less evident fact is
that Eqs. ({\ref{g-gic-nA'}) give the same $\vec P^{gic}_g$, $\vec J^{gic}_g$
as in Eqs. ({\ref{g-gic-nA'}). Here we give some detail, with the sofar suppressed
color indices added explicitly:
\begin{eqnarray}
&&\hat E_{phys} ^{ia} \vec \partial
\hat A_{phys}^{ib}= 2{\rm Tr}\left\{ \hat E_{phys} ^{ia} T^a \vec \partial
\hat A_{phys}^{ib} T^b \right\}\nonumber\\
&=&2{\rm Tr}\left\{ e^{-ig\Lambda} E ^{ia} T^a e^{ig\Lambda}
\vec \partial
(e^{-ig\Lambda} A_{phys}^{ib} T^b  e^{ig\Lambda}) \right\}\nonumber \\
&=& 2{\rm Tr}\left\{ e^{-ig\Lambda} E ^{ia} T^a e^{ig\Lambda}
\left((\vec \partial
e^{-ig\Lambda} )A_{phys}^{ib} T^b  e^{ig\Lambda} \right.\right.\nonumber \\
&&\left.\left.+ e^{-ig\Lambda} (\vec \partial A_{phys}^{ib} T^b)  e^{ig\Lambda}
+ A_{phys}^{ib} T^b (\vec \partial  e^{ig\Lambda})\right)\right\} \nonumber\\
&=& 2{\rm Tr}\left\{ E ^{ia} T^a \left((-ig \vec  A_{pure}^c T^c)A_{phys}^{ib} T^b
\right.\right. \nonumber  \\
&&+ \left.\left.\vec \partial A_{phys}^{ib} T^b
+A_{phys}^{ib} T^b (ig \vec A_{pure}^c T^c)\right) \right\} \nonumber\\
&=&2{\rm Tr}\left\{ E ^{ia} T^a \left(\vec \partial A_{phys}^{ib} T^b
-ig [\vec A_{pure}^c T^c, A_{phys}^{ib} T^b]\right)\right\}\nonumber \\
&=& 2{\rm Tr}\left\{ E ^{ia} T^a \vec {\cal D}_{pure} A_{phys}^{ib} T^b \right\}.
\end{eqnarray}
The proof for $\vec J^{gic}_g$ is very similar.

In the physical gauge with $\vec A_{pure}=0$, we have simultaneously
$\vec {\hat E}_{phys}=\vec E$ and $\vec {\hat A}_{phys}=\vec A_{phys}= \vec A$.
Then, a comparison of Eqs. (\ref{g-gic-nA'}) with Eqs. (\ref{nc}) reveals that
$\vec P^{gic}_g$ and $\vec J^{gic}_g$ are the translation and rotation
generators for the {\it gauge-invariant physical} gluon field $\vec {\hat A}_{phys}$.
Together with our earlier illustration for the quark sector, we see that, analogously
to the Abelian case, {\it $\vec P^{gic}_{q/g}$ and $\vec J^{gic}_{q/g}$ qualify as
the momentum and angular momentum operators for the gauge-invariant
physical quark/gluon field.}

{\it Summary.}---In gauge theories, the combination of
Poincar\'e subalgebra with gauge-invariance dictates a unique separation
of the translation and rotation generators (namely, the momentum and
angular momentum operators), as given by Eqs. (\ref{gic}) for the
Abelian case and Eqs. (\ref{gic-nA}) for the non-Abelian case.
An important further observation is that the matter-field and gauge-field parts
act as the translation and rotation generators for the gauge-invariant
physical component of the matter and gauge fields, respectively, and thus are
pertinent representation of the momentum and angular momentum of the physical
fields. In the naive canonical separation, on the other hand, the free-form
operators represent the momentum and angular momentum of the full matter field
or gauge field which also contain nonphysical gauge degrees of freedom.
Such operators are thus naturally gauge-dependent. We clarify that when going
from operators to matrix elements, the gauge-dependence in the naive canonical
separation relaxes to certain extent, but might not completely. In perturbative
calculations with quark and gluon states, the relaxation is observed 
only within a gauge class like the covariant gauge, but not from one
gauge class to another \cite{Chen09,Hood99}. 
The issue remains open, though, for a non-perturbative 
verification with color-singlet states, which presently can only be done on the 
lattice, and is encouraged. 

This work is supported by the China NSF via Grants No.
10875082 and No. 11035003, and by the NCET Program of the China
Ministry of Education. We also thank the [DOE's] Institute for Nuclear 
Theory at the University of Washington for its hospitality and the DOE 
for partial support during the completion of this work.

\end{document}